\documentclass[reprint,superscriptaddress,showpacs,amsmath,amssymb,nofootinbib,aps]{revtex4-1}
\usepackage{graphicx}
\usepackage{dcolumn}
\usepackage{bm}
\usepackage{amsthm,amsmath,amssymb}
\usepackage{mathrsfs}
\usepackage{multirow}
\usepackage[T1]{fontenc}
\usepackage{blindtext}
\usepackage[
colorlinks=true,        
citecolor=blue,         
linkcolor=blue,         
urlcolor=blue           
]{hyperref}             
\usepackage{float}

\begin{document}
\preprint{APS/123-QED}

\title{Testing supermassive primordial black holes with lensing signals of binary black hole merges}

\author{Huan Zhou}
\affiliation{School of Physics and Optoelectronic Engineering, Yangtze University, Jingzhou, 434023, China}

\author{Bin Liu}
\affiliation{Institute for Gravitational Wave Astronomy, Henan Academy of Sciences, Zhengzhou 450046, China}

\author{Zhengxiang Li}
\email{zxli918@bnu.edu.cn}
\affiliation{School of Physics and Astronomy, Beijing Normal University, Beijing 100875, China}
\affiliation{Institute for Frontiers in Astronomy and Astrophysics, Beijing Normal University, Beijing 102206, China}

\author{Xi-Jing Wang}
\affiliation{School of Physics and Technology, Wuhan University, Wuhan 430072, China}

\author{Kai Liao}
\affiliation{School of Physics and Technology, Wuhan University, Wuhan 430072, China}
\date{\today}

\begin{abstract}
Next-generation ground-based gravitational wave (GW) detectors are expected to observe millions of binary black hole mergers, a fraction of which will be strongly lensed by intervening galaxies or clusters, producing multiple images with characteristic distribution of time delay. Importantly, the predicted rate and properties of such events are sensitive to the abundance and distribution of strong lensing objects which directly depends on cosmological models. One such scenario posits the existence of supermassive primordial black holes (SMPBHs) in the early universe, which would enhance the formation of dark matter halos. This mechanism has been proposed to explain the abundance of high-redshift galaxies observed by James Webb Space Telescope. Crucially, the same cosmological model with SMPBHs would also leave a distinct imprint on the population of strongly lensed GWs. It predicts both an increased event rate and a modified distribution of time delays between the multiple images. Therefore, we propose statistical measurements of the rate and time delay distribution of strong lensing GW events as a powerful probe to directly constrain the abundance of SMPBHs. Considering $\Lambda$CDM cosmology with (non-)clustered SMPBHs, we find that the abundance of SMPBHs $f_{\rm PBH}$ with masses above $10^8~M_{\odot}$ is constrained to be $\sim10^{-4}$ at $95\%$ confidence level. It will be comparable and complementary to the currently available constraint from large scale structure observations.
\end{abstract}
\maketitle

\section{Introduction}
Since the first discovery of gravitational wave (GW) from binary black hole (BBH) merger GW150914~\cite{LIGO2016}, subsequent observing runs by the LIGO-Virgo-KAGRA collaboration have cataloged to several hundred compact binary coalescences, progressively expanding our sample through the GWTC series~\cite{LIGOScientific:2018mvr, LIGOScientific:2020ibl, LIGOScientific:2021usb, KAGRA:2021vkt, LIGOScientific:2025pvj}. 
The rapidly growing dataset from these observations has firmly established GWs as a unique and powerful probe for addressing fundamental questions in astrophysics and cosmology, e.g. offering new measurements of the Hubble constant $H_0$ without relying on the distance ladder~\cite{LIGOScientific:2017adf, LIGOScientific:2019zcs,Chen:2017rfc, DES:2019ccw,Hotokezaka:2018dfi, LIGOScientific:2018gmd,
Feeney:2020kxk, Ezquiaga:2022zkx, Gray:2021sew, Palmese:2021mjm, LIGOScientific:2021aug, Shiralilou:2022urk, Mukherjee:2022afz, Huang:2022rdg}, probing the nature of particle dark matter~\cite{Arvanitaki:2014wva, Baryakhtar:2017ngi, Bertone:2019irm, Kavanagh:2020cfn, Kadota:2023wlm}, probing the nature of compact dark matter from micro-lensing effect of GWs ~\cite{Jung2019, Liao2020,Basak2021,Urrutia2021,Wang2021,Zhou2022, Guo2022, Urrutia2023,Fairbairn2022, LIGO2023, GilChoi2023, Barsode:2024wda, Cheung2024}, precise probes of cosmological parameters from strong lensing effect of GWs~\cite{Liao:2017ioi,Wei:2017emo, Li:2019rns, Yang:2018bdf, Liu:2019dds,Jana:2022shb,Jana:2024uta,Jana:2024dhc,Ying:2025eem, Chen:2025xeg, Maity:2025apt}, and unique and extremely precise tests of General Relativity (GR)~\cite{LIGOScientific:2016lio,LIGOScientific:2017zic, LIGOScientific:2019fpa,LIGOScientific:2020tif}. The imminent advent of next-generation ground-based GW detectors, i.e. Einstein Telescope (ET)~\cite{Punturo:2010zza} and Cosmic Explorer (CE)~\cite{Reitze:2019iox}, promises a quantum leap in sensitivity, forecasting the detection of millions of BBH merged events annually up to high redshift ($z\geq10$)~\cite{Hall:2019xmm}, thereby unlock unprecedented precision across multiple scientific domains.

The existence of dark matter, which is supported by multiple lines of observational evidence, permits candidate masses across an extremely wide range—from ultralight particles to supermassive primordial black holes (SMPBHs)~\cite{Sasaki2018,Green:2020jor,Carr:2020gox,Carr:2021bzv}. Primordial black holes (PBHs)~\cite{Hawking:1971ei,Carr:1974nx,Carr:1975qj}, which formed in the early universe and are considered a potential component of dark matter, exhibit various mass window—spanning from scale on Planck mass ($10^{-5}~{\rm g}$) to supermassive black holes (SMBHs) in galactic centers. Over several decades, intensive observational searches for PBHs have been conducted. These efforts have produced a variety of methods to constrain the PBH abundance, i.e. commonly quantified as the fraction of PBH in dark matter ($f_{\mathrm{PBH}} = \Omega_{\mathrm{PBH}} / \Omega_{\mathrm{DM}}$), across different mass windows utilizing both direct and indirect constraints (see reviews in~\cite{Sasaki2018,Green:2020jor,Carr:2020gox,Carr:2021bzv}). Direct constraints on PBHs are obtained from observational signatures of their intrinsic gravitational influence, and are therefore independent of PBH formation mechanisms~\cite{Sasaki2018}. These constraints are primarily classified into four categories based on the manners PBHs affect: gravitational lensing~\cite{Nemiroff:2001bp,Wilkinson:2001vv,Munoz:2016tmg, Zumalacarregui:2017qqd, Niikura:2017zjd, Jung2019, CHIMEFRB:2022xzl}, dynamical effects~\cite{Brandt:2016aco,Koushiappas:2017chw}, the effect of accretion~\cite{Ali-Haimoud:2016mbv, Aloni:2016kuh}, and impacts on large-scale structure growth~\cite{Carr:2018rid,Murgia:2019duy}. Indirect constraints arise not from PBHs themselves but from observables strongly linked to PBH scenarios~\cite{Sasaki2018}. While such constraints do not apply to all possible PBH models, they remain effective in ruling out specific formation channels or parameter regimes, such as null detection of scalar-induced GW measured by pulsar timing array (PTA) experiment~\cite{Chen:2019xse} and cosmic microwave background (CMB) spectral distortions from the primordial density perturbations~\cite{Carr:1993aq,Carr:1994ar}. The direct detection of GWs by laser-interferometer observatories represents a fundamentally novel approach to search for PBHs, independent of electromagnetic signature. For instance, the detection of GW bursts from BBH mergers serves as one of the most promising ways to constrain PBH population information~\cite{Wu:2020drm, DeLuca:2020qqa, DeLuca:2021wjr, Hutsi:2020sol, Ng:2022agi,Franciolini:2022tfm,Chen:2022fda}, the non-detection of a predicted stochastic GW background can place stringent upper limits on abundance of PBH~\cite{Wang:2016ana,DeLuca:2020qqa,Hutsi:2020sol}.

Strong lensing GWs provide a unique and complementary probe of cosmology. It can produce multiple images of a single GW signal, arriving at the detector with measurable time delays. During their operational lifetime, next-generation ground-based GW detectors are expected to detect more than thousands of such strongly lensed events by galaxies and clusters~\cite{Oguri:2018muv,Yang:2021viz,Smith:2022vbp,Jana:2022shb,Jana:2024uta}. The precise measurement of the time delay distribution between these lensed images offers a powerful probe for cosmological studies~\cite{Jana:2022shb,Jana:2024uta,Jana:2024dhc, Ying:2025eem,Maity:2025apt}. Recent James Webb Space Telescope (JWST) observations have revealed numerous quasars powered by SMBHs existing as early as the first few hundred million years after the Big Bang~\cite{CEERSTeam:2023qgy,Goulding:2023gqa,Maiolino:2023bpi,Maiolino:2023zdu, Bogdan:2023ilu, Natarajan:2023rxq, Kovacs:2024zfh}. The presence of these SMBHs at high redshifts presents a major theoretical challenge, which has prompted the proliferation of models focused on early SMPBH seeding mechanisms~\cite{Liu:2022bvr, Hutsi:2022fzw, Gouttenoire:2023nzr,Huang:2024aog,Matteri:2025klg}. Incorporating SMPBHs into $\Lambda$CDM cosmology would elevate the halo mass function, thereby promoting the formation of massive galaxies at various redshift bins. This enhancement is predicted to generate observable deviations from the standard $\Lambda$CDM cosmology, specifically in the future-detected population of strong lensing GWs and their time delay distribution. In this work, we propose a method, based on the $\Lambda$CDM cosmology and incorporating both clustered and non-clustered (Poisson) SMPBH, to directly constrain the abundance of SMPBHs by exploiting these predicted observational signatures from strong lensing GW events. 

This paper is organized as follows: In Section~\ref{sec2}, we demonstrate how SMPBHs modify the halo mass function by influencing the matter power spectrum. In Section~\ref{sec3}, we present the probability of strong lensing GWs and time delay distribution. Results for constraining the abundance of SMPBHs presented in Section~\ref{sec4}. Finally, Section~\ref{sec5} presents conclusion and discussion. In this paper, we adopt the concordance $\Lambda$CDM cosmological model with the best-fitting parameters from the recent Planck observations~\cite{Planck2018}, and the natural units of $G=c=1$ in all equations.

\begin{figure*}
    \centering
    \includegraphics[width=0.99\textwidth, height=0.3\textwidth]{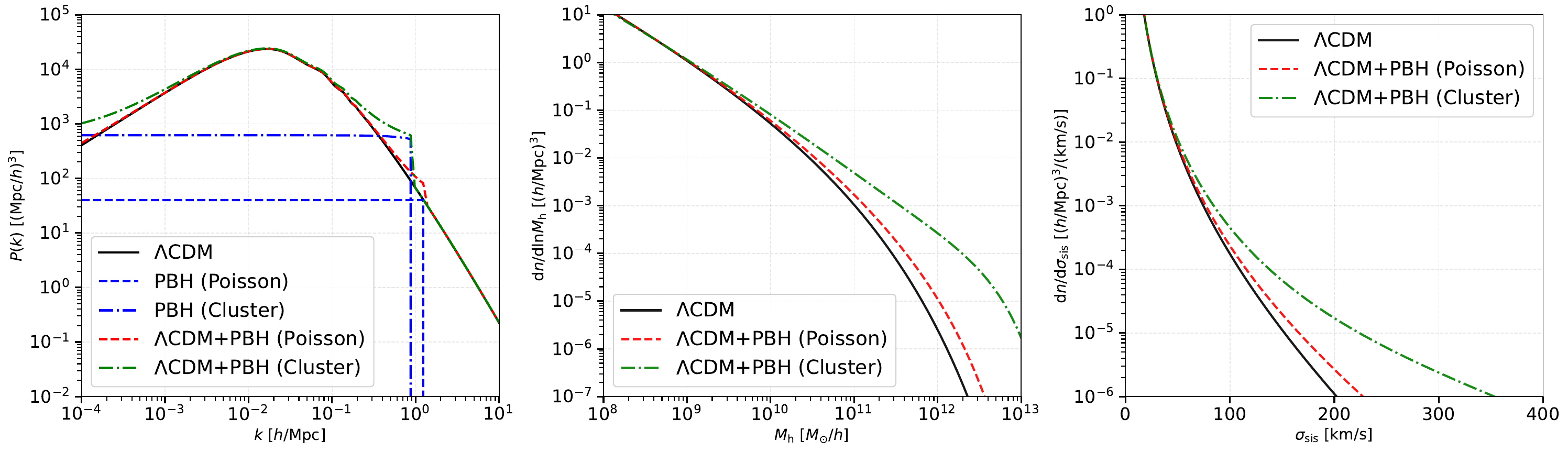}
     \caption{{\bf Left:} The linear matter power spectrum at $z=0$ for $\Lambda$CDM cosmology (black solid); for $\Lambda$CDM+PBH cosmology with initially Poisson distributed SMPBHs ($\xi_0=0$, red dashed); and for initially clustered SMPBHs ($x_{\rm cl}=1~{\rm Mpc}$, $\xi_0=10$, green dash-dotted) respectively, where $M_{\rm PBH}=10^9~M_{\odot}$, $f_{\rm PBH}=10^{-3}$. The blue dashed and dash-dotted lines represent the isocurvature perturbations originating from the Poisson distributed and clustered SMPBHs with cutoff scale $k_{\rm cut}$, respectively.  {\bf Middle:} Corresponding halo mass function for $\Lambda$CDM and $\Lambda$CDM+PBH cosmology at redshift $z=7$. {\bf Right:} Halo mass function as function of the velocity dispersion of lenses in singular isothermal sphere model.}\label{fig1}
\end{figure*}

\section{Effects of SMPBH}\label{sec2}
\subsection{Power Spectrum with SMPBH}\label{sec2-1}

\begin{figure*}
    \centering
    \includegraphics[width=0.9\textwidth, height=0.36\textwidth]{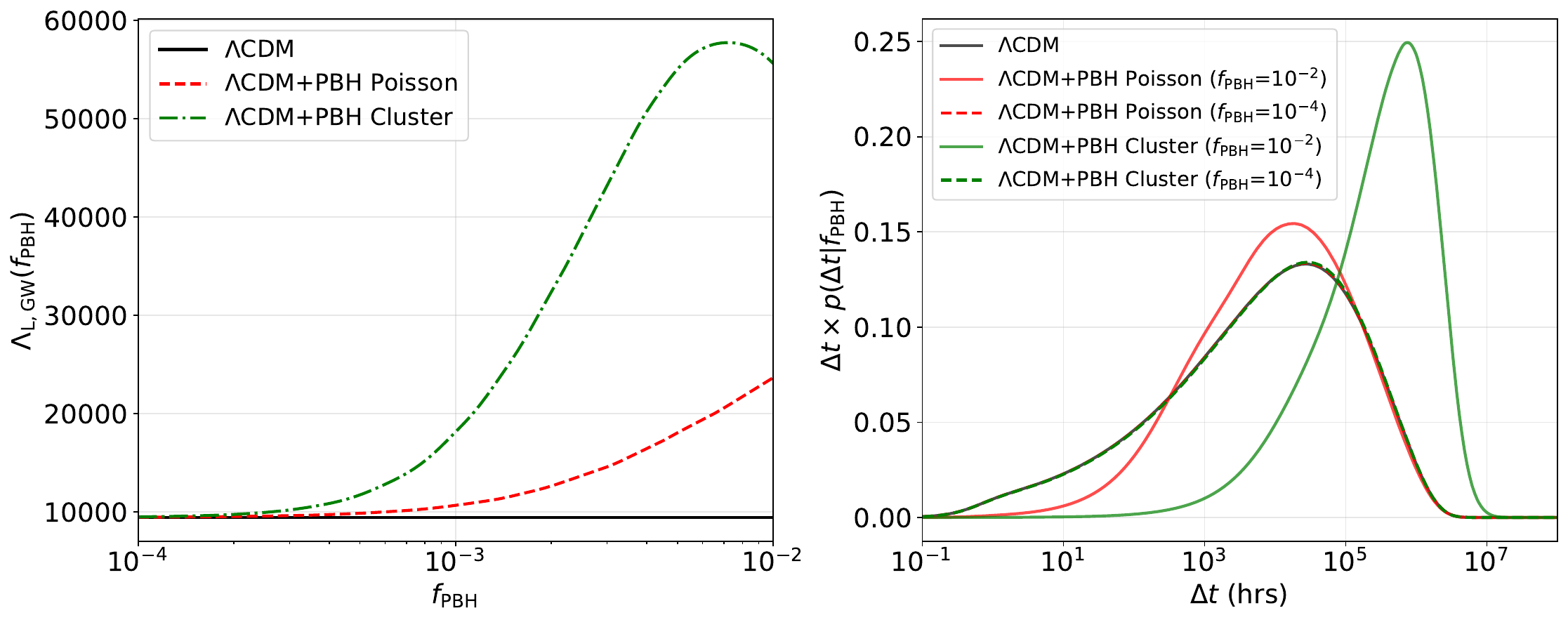}
     \caption{{\bf Left:} Expected number of strong lensing GW events $\Lambda_{\rm L,GW}(f_{\rm PBH})$ for both of $\Lambda$CDM and $\Lambda$CDM+PBH cosmology ($x_{\rm cl}=1~{\rm Mpc}$, $\xi_0=[0,10]$, $M_{\rm PBH}=10^9~M_{\odot}$), assuming a merger rate $R=5\times10^5~{\rm yr^{-1}}$ and observation duration $T_{\rm obs}=10~{\rm yrs}$. {\bf Right:} The strong lensing time delay distributions $p(\Delta t|f_{\rm PBH})$ at different values of $f_{\rm PBH}$ for different cosmology.}\label{fig2}
\end{figure*}

Because PBHs can be regarded as discrete objects, the clustering of PBHs, which extends beyond Poisson distribution, is characterized by their correlation functions. While the precise form depends on the specific model, we adopt a simplified representation for the PBH correlation function with monochromatic mass spectrum, following the approach established in previous papers written as~\cite{DeLuca:2022uvz,Huang:2024aog, Zhang:2025tgm}
\begin{equation}\label{eq2-1}
\xi_{\rm cl}(x) =
\begin{cases}
\xi_0, & x \leq x_{\rm cl}, \\
0, & \text{otherwise},
\end{cases}
\end{equation}
where $x=|\vec{x}|$ is the comoving scale, and $x_{\rm cl}$ is the comoving clustering scale. The density fluctuation of PBHs is 
\begin{equation}\label{eq2-2}
\delta_{\rm PBH}(\vec{x})=\frac{\rho_{\rm PBH}}{\bar{\rho}_{\rm PBH}}-1=\frac{1}{\bar{n}_{\rm PBH}}\sum\limits_{i}\delta_{\rm D}(\vec{x}-\vec{x}_i)-1
\end{equation}
where $\delta_{\rm D}(\vec{x})$ is the three-dimensional Dirac distribution, and $\bar{n}_{\rm PBH}$ is average comoving number density of PBHs for monochromatic mass function 
\begin{equation}\label{eq2-3}
\bar{n}_{\rm PBH}=\frac{f_{\rm PBH}\Omega_{\rm DM}\rho_{\rm c}}{M_{\rm PBH}}\approx10^{11}f_{\rm PBH}\frac{M_{\odot}}{M_{\rm PBH}}~{\rm (h/Mpc)^{3}}
\end{equation}
where $\rho_{\rm c}$ is the critical density of universe, and $\Omega_{\rm DM}$ is the dark matter density parameter at the present universe. The two-point correlation function of PBHs is contributed by two components as
\begin{equation}\label{eq2-4}
\xi_{\rm PBH}(\vec{x})=\xi_{\rm Poisson}(\vec{x})+\xi_{\rm Cluster}(\vec{x})=\frac{\delta_{D}(\vec{x})}{\bar{n}_{\rm PBH}}+\xi_{\rm cl}(x).
\end{equation}
The power spectrum of the density perturbations of PBHs from Fourier transform of the two-point correlation function $\xi_{\rm PBH}(\vec{x})$ as
\begin{equation}\label{eq2-5}
\begin{split}
&P_{\rm PBH}(k) = \int d^3\vec{x}e^{-i\vec{k}\cdot\vec{x}}\xi_{\rm PBH}(\vec{x})\\
&=P_{\rm Poisson}(k)+P_{\rm Cluster}(k),
\end{split}
\end{equation}
where $k=|\vec{k}|$, $P_{\rm Poisson}(k)$ is independent of scale $k$ and is from the Poisson fluctuation induced by the fluctuation of the number of PBHs, as
\begin{equation}\label{eq2-6}
P_{\rm Poisson}(k) = \frac{1}{\bar{n}_{\rm PBH}}=\frac{M_{\rm PBH}}{f_{\rm PBH}\Omega_{\rm DM}\rho_{\rm c}}.
\end{equation}
In addition, $P_{\rm Cluster}(k)$ comes from the clustering distribution of PBHs
\begin{equation}\label{eq2-7}
P_{\rm Cluster}(k) = \frac{4\pi\xi_0x_{\rm cl}^3[\sin(r_{\rm cl})-r_{\rm cl}\cos(r_{\rm cl})]}{r_{\rm cl}^3},
\end{equation}
where $r_{\rm cl}\equiv kx_{\rm cl}$. The isocurvature perturbations induced by PBHs through $f_{\rm PBH}^2P_{\rm PBH}(k)$ grow during the matter-dominated era and are given by ($z=0$)
\begin{equation}\label{eq2-8}
P_{\rm iso}(k)=
\begin{cases}
[f_{\rm PBH}D_{\rm PBH}(0)]^2P_{\rm PBH}(k), & k \leq k_{\rm cut}, \\
0, & \text{otherwise},
\end{cases}
\end{equation}
where $D_{\rm PBH}(0)\simeq(1+\frac{3\gamma}{2a_{-}}(1+z_{\rm eq}))^{a_{-}}$ ($\gamma=\Omega_{\rm DM}/\Omega_{\rm m}$, $z_{\rm eq}\approx3400$, $a_{-}=(\sqrt{1+24\gamma}-1)/4$)) is  the growth factor of isocurvature perturbations~\cite{Inman:2019wvr}, and isocurvature term is truncated at the scale $k_{\rm cut}$ where we expect the linear Press-Schechter theory to break down at small scales. The order of the cutoff scale of the spectrum is the inverse mean separation between PBHs (or clusters)~\cite{Inman:2019wvr, DeLuca:2020jug, Hutsi:2022fzw, Gouttenoire:2023nzr, Huang:2024aog}
\begin{equation}\label{eq2-9}
k_{\rm cut} =
\begin{cases}
(2\pi^2\bar{n}_{\rm PBH})^{1/3}, & \xi_0=0, \\
(2\pi^2\bar{n}_{\rm cl})^{1/3}, & \text{otherwise},
\end{cases}
\end{equation}
where $\bar{n}_{\rm cl}$ is 
\begin{equation}\label{eq2-10}
\bar{n}_{\rm cl}=\frac{\bar{n}_{\rm PBH}}{N_{\rm cl}}~~~\bigg(N_{\rm cl}=1+\bar{n}_{\rm PBH}\int d^3\vec{x}\xi_{\rm cl}(x)\bigg),
\end{equation}
where $N_{\rm cl}$ is the average number of PBHs in a cluster. As shown in Equation~(\ref{eq2-8}), the Poisson power spectrum $P_{\rm iso,Poisson}$ is proportional to $f_{\rm PBH}M_{\rm PBH}$, whereas the clustering power spectrum $P_{\rm iso,Cluster}$ scales as $f_{\rm PBH}^2\xi_0x_{\rm cl}^3$ and is independent of the PBH mass $M_{\rm PBH}$ in the regime $P_{\rm iso,Cluster}\gg P_{\rm iso,Poisson}$. The total power spectrum of the $\Lambda$CDM+PBH cosmology $P_{\rm tot}(k)$ including PBH isocurvature perturbations
and $\Lambda$CDM standard adiabatic model can be expressed as
\begin{equation}\label{eq2-11}
P_{\rm tot}(k)= P_{\rm ad}(k)+P_{\rm iso}(k).
\end{equation}
In the left panel of Figure~\ref{fig1}, we show the total and PBH isocurvature perturbations power spectrum in the $\Lambda$CDM(+PBH) cosmology for initially (non)clustered SMPBHs with $f_{\rm PBH}=10^{-3}$, $M_{\rm PBH}=10^{9}~M_{\odot}$ at $z=0$. It is clear that the clustered PBH model is truncated at larger scales, and it contributes more to the power spectrum on large scales.

\begin{figure*}
    \centering
    \includegraphics[width=0.9\textwidth, height=0.36\textwidth]{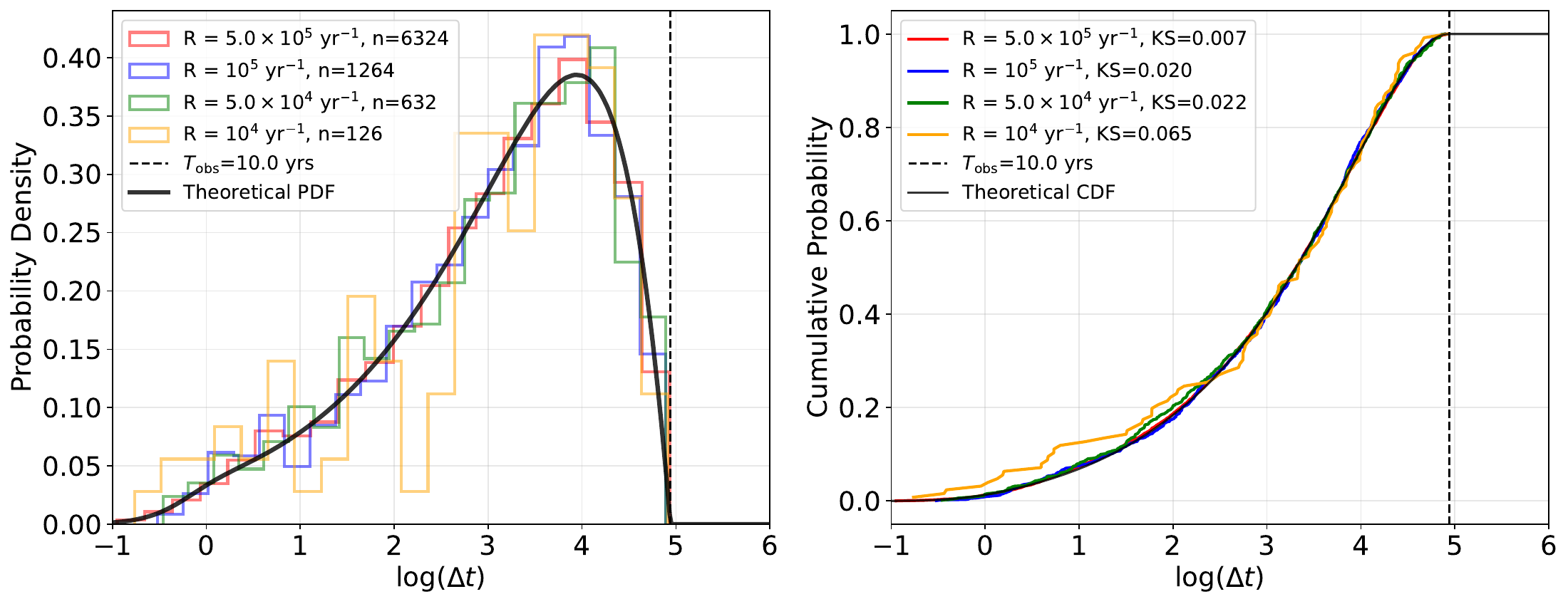}
    \includegraphics[width=0.9\textwidth, height=0.36\textwidth]{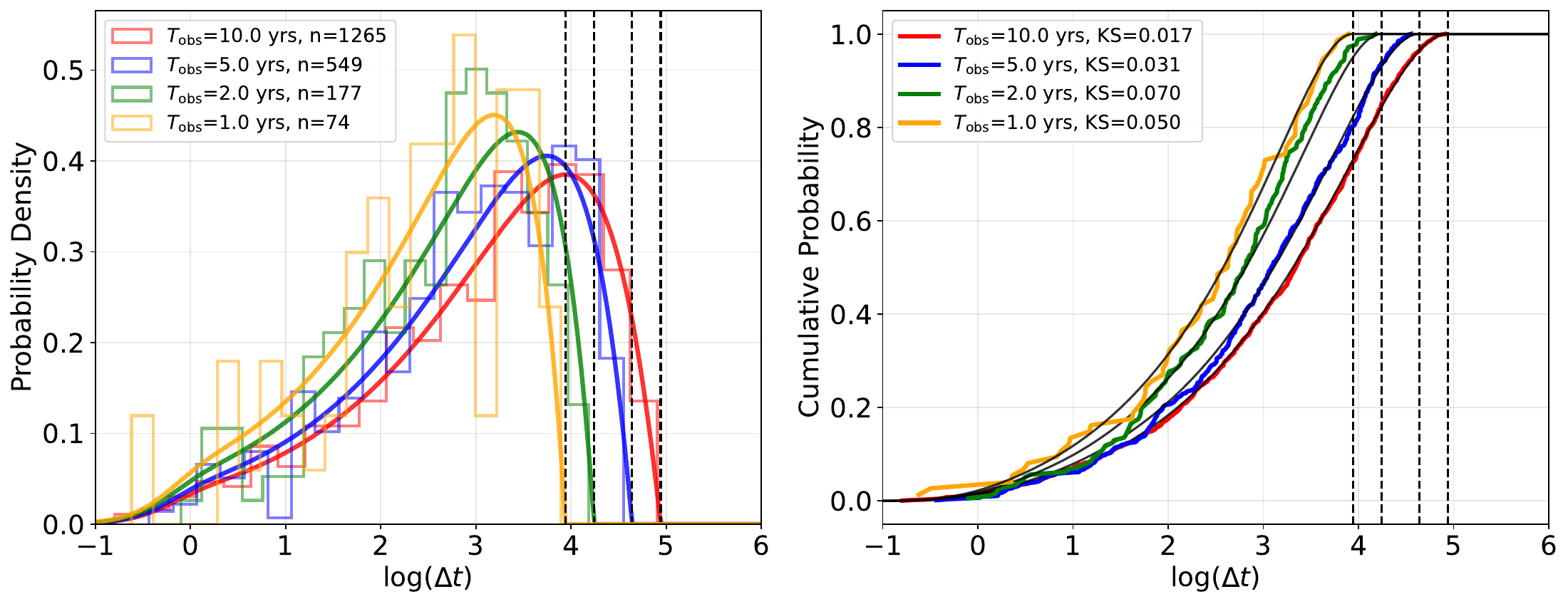}
     \caption{We simulate the population of observable strong lensing GW events based on the $\Lambda$CDM cosmology as fiducial model, considering different BBH detection rates and observing times $T_{\rm obs}$.
     {\bf Upper Left:} Detectable time delay distributions $p(\Delta t|f_{\rm PBH},T_{\rm obs})$ and corresponding event counts $\Lambda_{\rm L,GW}(f_{\rm PBH},T_{\rm obs})$ for a fixed observational duration $T_{\rm obs} = 10~{\rm yrs}$ under four different BBH detection rates $R \in[5 \times 10^5,\; 1 \times 10^5,\; 5 \times 10^4,\; 1 \times 10^4]~\rm{yr^{-1}}$. Here, $n$ denotes the realized number of observed lensed events $N_{\rm obs}$ in each mock sample. 
     {\bf Upper Right:} Comparison between the theoretical cumulative distribution functions (CDFs; black solid curve) and the sample-derived CDFs for the corresponding cases. The maximum vertical distance between each pair of CDFs is quantified by the Kolmogorov-Smirnov (K-S) statistic $D$ indicated in the plot.
     {\bf Lower:} Similar to the upper panel but with the detection rate fixed at $R = 1\times10^5~\rm{yr^{-1}}$ and for varying observational durations $T_{\rm obs}\in[10,\;5,\;2,\;1]~\rm{yrs}$.
     }\label{fig3}
\end{figure*}

\subsection{Halo Mass Function }\label{sec2-2}
Under the assumption of linear perturbation theory and Gaussian density fluctuations，the comoving number density of halos is given by  Press-Schechter formalism~\cite{Press:1973iz}
\begin{equation}\label{eq3-1}
\frac{dn_{\rm h}}{d\ln M_{\rm h}}=\frac{\rho_{\rm m}}{M_{\rm h}}\frac{d\ln \sigma^{-1}}{d\ln M_{\rm h}}f(\sigma),
\end{equation}
where $\rho_{\rm m}$ is the average density of matter, $f(\sigma)$ represents the fraction of mass that has collapsed to form halos, and $\sigma$ is the root-mean-square of the matter density fluctuation through a convolution of the total matter power spectrum with a spherical top-hat smoothing Kernel of radius,
\begin{equation}\label{eq3-2}
\sigma^2(M_{\rm h})=\int P_{\rm tot}(k)W^2(kR_{\rm h})\frac{k^2dk}{2\pi^2},
\end{equation}
where $W(kR_{\rm h})$ is the Fourier transform of the real-space
top-hat window function as
\begin{equation}\label{eq3-3}
W(kR_{\rm h})=\frac{3[\sin(kR_{\rm h})-kR_{\rm h}\cos(kR_{\rm h})]}{(kR_{\rm h})^3},
\end{equation}
$R_{\rm h}=(\frac{3M_{\rm h}}{4\pi\rho_{\rm m}})^{1/3}$ is the comoving radius associated with the halo mass window $M_{\rm h}$.
In Equation~(\ref{eq3-1}), we utilize the Sheth-Tormen halo mass
function~\cite{Sheth:1999mn,Sheth:1999su}
\begin{equation}\label{eq3-4}
\begin{split}
&f(\sigma)=A(p)\sqrt{\frac{2q}{\pi}}\bigg[1+\bigg(\frac{\sigma^2}{q\delta^2_{\rm c}(z)}\bigg)^p\bigg]\times\\
&\frac{\delta_{\rm c}(z)}{\sigma}\exp\bigg(-\frac{q\delta^2_{\rm c}(z)^2}{2\sigma}\bigg),
\end{split}
\end{equation}
with the fitting coefficients $A(p)=0.3222$, $q=0.707$, $p=0.3$, and the critical collapse density $\delta_c(z)=1.686/D(z)$ ($D(z)$ is the linear growth function, normalized so that $D(0)=1$).

The halo mass functions as a function of lens mass and velocity dispersion, respectively are presented in the middle and right panels of Figure~\ref{fig1} respectively, under the standard $\Lambda$CDM model and the $\Lambda$CDM+PBH scenarios~\footnote{Due to the strong degeneracy among $f_{\rm PBH}$, $\xi_0$, and $x_{\mathrm{cl}}$, excessive clustering  would lead to significant SMPBH mergers~\cite{DeLuca:2022uvz,Huang:2024aog} and conflict with large-scale CMB observational constraints~\cite{Planck:2018jri}, we set the parameter $\xi_0 = [0,10]$ and $x_{\mathrm{cl}} = [1,10]~\mathrm{Mpc}$ in the following analysis.}. The presence of SMPBHs enhances the matter power spectrum. This enhancement shifts to larger scales and its amplitude increases if the SMPBHs are initially clustered (compared to the Poisson scenario), resulting in an amplified halo mass function for heavier halos.

\begin{figure*}
    \centering
    \includegraphics[width=0.9\textwidth, height=0.36\textwidth]{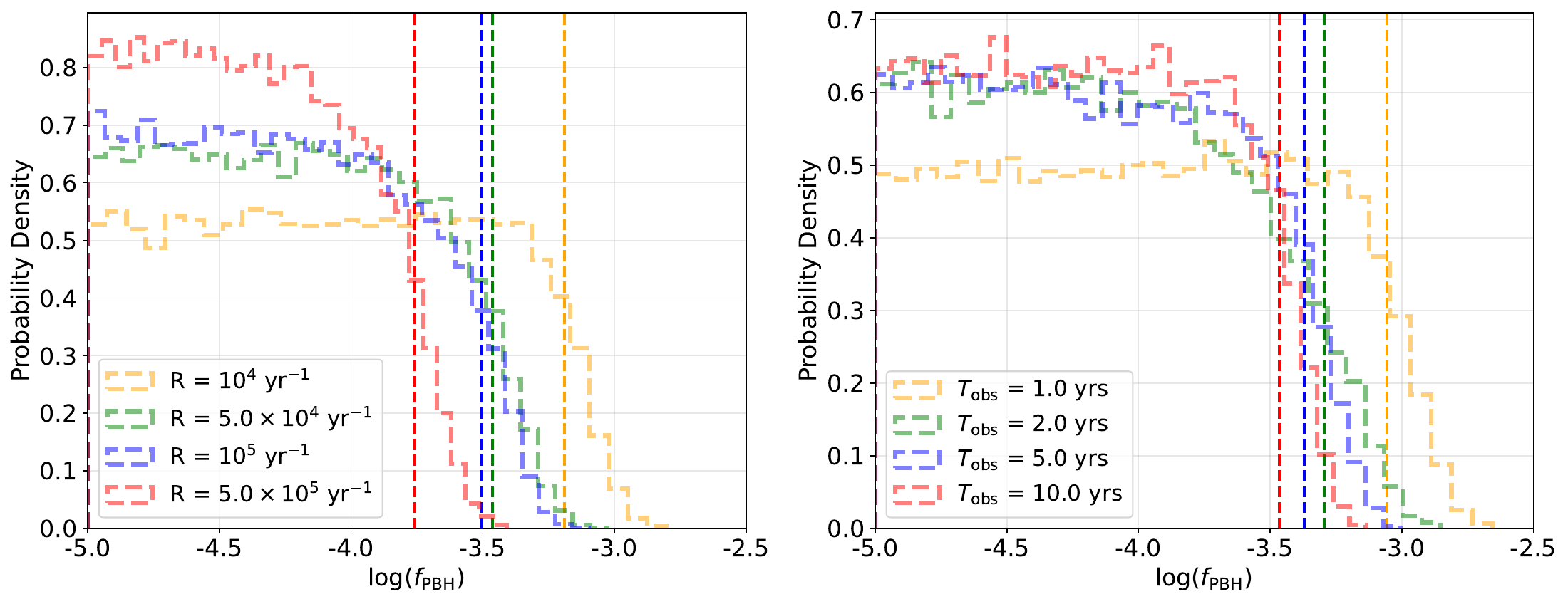}
    \includegraphics[width=0.9\textwidth, height=0.36\textwidth]{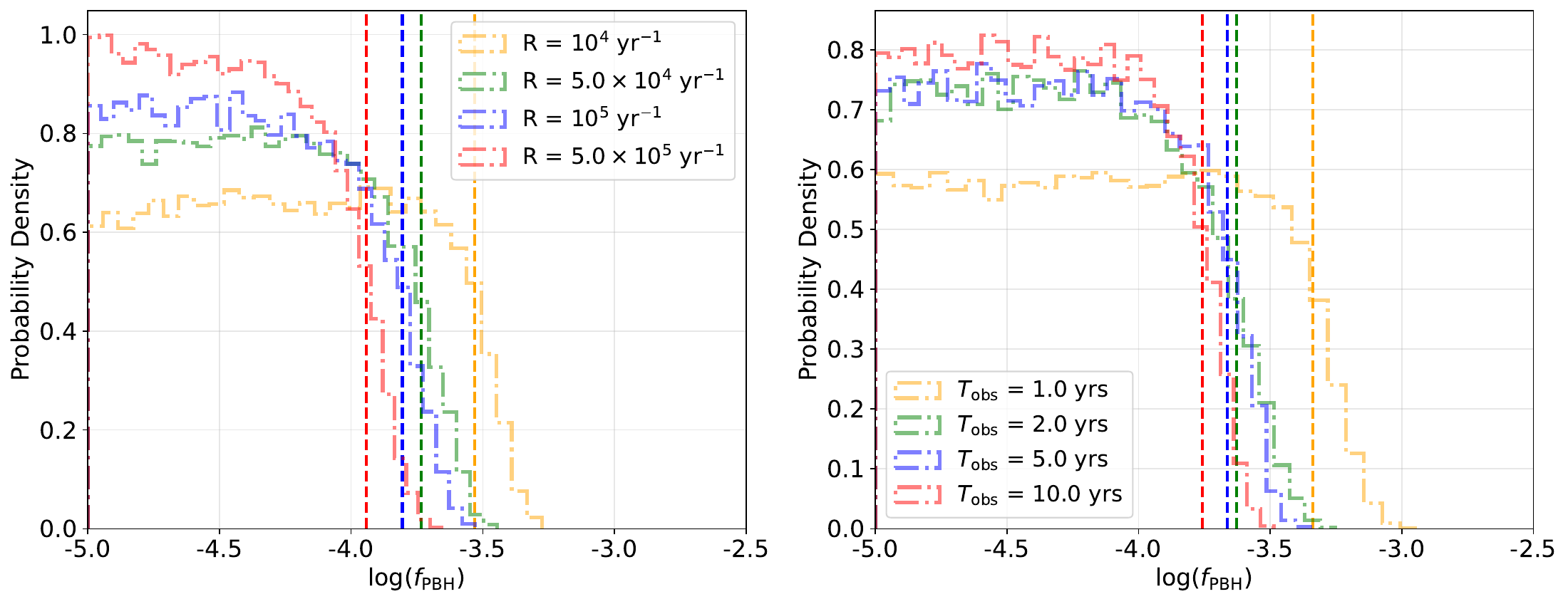}
     \caption{Constraints on $f_{\rm{PBH}}$ at $95\%$ confidence level in $\Lambda$CDM+PBH cosmology with $M_{\rm{PBH}} = 10^9~M_{\odot}$.
     {\bf Upper Left:} Initially Poisson distributed SMPBHs (i.e., $\xi_0 = 0$) under different BBH detection rates $R\in[5\times10^5, 1\times10^5,5\times10^4,1\times10^4]~{\rm yr^{-1}}$, for a fixed observational duration $T_{\rm obs}=10~{\rm yrs}$.
     {\bf Upper Right:} The same Poisson-distributed scenario, but with fixed detection rate $R=1\times10^5~{\rm yr^{-1}}$ and varying observational durations $T_{\rm obs}\in[10,\;5,\;2,\;1]~\rm{yrs}$.
     {\bf Lower Left:} Initially clustered SMPBHs ($x_{\mathrm{cl}} = 1~\rm{Mpc}$, $\xi_0 = 10$) under the same set of detection rates $R$ as in the upper left panel.
     {\bf Lower Right:} The same clustered scenario, under the same varying $T_{\mathrm{obs}}$ as in the upper right panel.
     }\label{fig4}
\end{figure*}

\section{Strong gravitational lensing probability and time delay}\label{sec3}
Given that the wavelengths of GWs from BBH are far smaller than both the lens scales and the cosmological distances, we can adopt the geometric optics and thin lens approximations. We further model the lenses as singular isothermal spheres (SIS) in the mass range $M_{\rm h}\in[10^{10},10^{15}]~M_{\odot}$, which provides a reasonable first-order description for galaxies located at the center of spherical dark matter halos. Under the assumption that selection effects can be neglected for BBH mergers at redshifts $z_{\rm s}<10$, the expected number of strong lensing GW events is estimated as
\begin{equation}\label{eq4-1}
\Lambda_{\rm L,GW}(f_{\rm PBH}) = RT_{\rm obs}\int_{0}^{z_{\max}}p(z_{\rm s})\tau(z_{\rm s}|f_{\rm PBH})dz_{\rm s}
\end{equation}
where $R$ and $T_{\rm obs}$ is BBH detection rate and observation period for next-generation ground-based GW detectors respectively, $p(z_{\rm s})$ is the redshift distribution of GW sources
\begin{equation}\label{eq4-2}
\begin{split}
p(z_{\rm s})=\frac{1}{N_{\rm s,GW}}\frac{dN_{\rm s,GW}(z_{\rm s})}{dz_{\rm s}}
\\
=\frac{1}{N_{\rm s,GW}}\frac{R_{\rm BBH}(z_{\rm s})}{1+z_{\rm s}}\frac{dV_{\rm c}}{dz_{\rm s}},
\end{split}
\end{equation}
where $R_{\rm BBH}(z_{\rm s})$ denotes the source-frame merger rate per unit comoving volume and incorporates a delay time distribution relative to the star formation rate~\cite{Mukherjee:2021qam,Urrutia2021}. $\tau(z_{\rm s}|f_{\rm PBH})$ is the lensing optical depth written as
\begin{equation}\label{eq4-3}
\begin{split}
\tau(z_{\rm s}|f_{\rm PBH})=\int_{0}^{z_{\rm s}}\int_{\sigma_{\rm sis,min}}^{\sigma_{\rm sis,max}}dz_{\rm l}d\sigma_{\rm sis}\frac{d\chi(z_{\rm l})}{dz_{\rm l}}\times\\
\frac{dn(\sigma_{\rm sis},z_{\rm l}|f_{\rm PBH})}{d\sigma_{\rm sis}}(1+z_{\rm l})^2\Sigma_{\rm L}(\sigma_{\rm sis},z_{\rm l},z_{\rm s}),
\end{split}
\end{equation}
where $\chi(z_{\rm l})$ is the comoving distance to the lens, $dn(\sigma_{\rm sis}, z_{\rm l}|f_{\rm PBH})/d\sigma_{\rm sis}$ denotes the comoving number density of halos as a function of velocity dispersion of lens, and velocity dispersion $\sigma_{\rm sis}$ is related to the halo mass through the virial velocity, allowing for mutual conversion between these quantities
\begin{equation}\label{eq4-4}
\begin{split}
\sigma_{\rm sis}\simeq\sqrt{\frac{M_{\rm h}}{R_{\rm vir}}},~~~M_{\rm h}=\frac{4\pi}{3}R_{\rm vir}^3\Delta_{\rm h}(z)\bar{\rho}(z),
\end{split}
\end{equation}
where $\Delta_{\rm h}(z)$ is the overdensity of the halo and $\bar{\rho}(z)$ is the mean density of the universe, both of which depend on redshift. In optical depth, $\Sigma_{\rm L}(\sigma_{\rm sis},z_{\rm l},z_{\rm s})$ is the scattering section of lens
\begin{equation}\label{eq4-5}
\Sigma_{\rm L}(\sigma_{\rm sis},z_{\rm l},z_{\rm s})=\pi\theta_{\rm E}^2D_{\rm l}^2y^2=16\pi^3\sigma_{\rm sis}^4\frac{D_{\rm ls}^2D_{l}^2}{D_{\rm s}^2}y^2.
\end{equation}
where $y$ quantifies the projected offset of the source on the lens plane, normalized by the Einstein radius $\theta_{\rm E}=4\pi\sigma_{\rm sis}^2\frac{D_{\rm ls}}{D_{\rm s}}$, the terms $D_{\rm l}$, $D_{\rm s}$, and $D_{\rm ls}$ denote the angular diameter distances corresponding to the lens, the source, and the lens-source separation, respectively. The left panel of Figure~\ref{fig2} presents the number of strong lensing GWs versus the abundance of SMPBHs $f_{\rm PBH}$. We find that in the $\Lambda$CDM cosmology, the strong lensing count is independent of $f_{\rm PBH}$. Whereas in the $\Lambda$CDM+PBH scenario, the clustered model yields a larger halo mass function than the Poisson model, leading to a higher lensing rate which grows substantially with $f_{\rm PBH}$.

In the SIS lens model, the strong lensing time delay $\Delta t$ between two images is
\begin{equation}\label{eq4-6}
\Delta t(\lambda|f_{\rm PBH})= 32\pi^2y\sigma_{\rm sis}^4(1+z_{\rm l})\frac{D_{\rm l}D_{\rm ls}}{D_{\rm s}}.
\end{equation}
The expected time delay distribution $p(\Delta t|f_{\rm PBH})$ by marginalizing the distribution of time delay over all other parameters $\lambda\equiv[y,\sigma_{\rm sis}, z_{\rm l},z_{\rm s}]$ is
\begin{equation}\label{eq4-7}
p(\Delta t|f_{\rm PBH})=\int p(\Delta t|\lambda,f_{\rm PBH})p(\lambda|f_{\rm PBH})d\lambda,
\end{equation}
where $p(\Delta t|\lambda,f_{\rm PBH})$ is the distribution of the time delay $\Delta t$ for given $\lambda$ and $f_{\rm PBH}$. Since the statistical error in measuring the arrival time of GWs ($\sim\rm ms$) could be negligible~\cite{Liao:2017ioi}, we can approximately represent $p(\Delta t|\lambda,f_{\rm PBH})$ as follows 
\begin{equation}\label{eq4-8}
\begin{split}
&p(\Delta t|\lambda,f_{\rm PBH})= \lim_{\sigma_{\Delta t}\rightarrow{0}}\frac{1}{\sigma_{\Delta t}\sqrt{2\pi}}\times\\
&\exp\left(-\frac{(\Delta t-\Delta t(\lambda|f_{\rm PBH}))^2}{2\sigma_{\Delta t}^2}\right)=\\
&\delta(\Delta t-\Delta t(\lambda|f_{\rm PBH})).
\end{split}
\end{equation}
$p(\lambda|f_{\rm PBH})$ is the distribution of $\lambda$ for given $f_{\rm PBH}$ obtained by optical depth theory and redshift distribution of GW sources
\begin{equation}\label{eq4-9}
p(\lambda|f_{\rm PBH})=p(y,z_{\rm l},\sigma_{\rm sis}|z_{\rm s},f_{\rm PBH})p(z_{\rm s}),
\end{equation}
where $p(y,z_{\rm l},\sigma_{\rm sis}|z_{\rm s},f_{\rm PBH})$ is calculated from the differential optical depth by
\begin{equation}\label{eq4-10}
\begin{split}
p(y,z_{\rm l},\sigma_{\rm sis}|z_{\rm s},f_{\rm PBH})=\frac{1}{\tau(z_{\rm s},f_{\rm PBH})}\frac{d\tau}{dydz_{\rm l}d\sigma_{\rm sis}}=\\
\frac{1}{\tau(z_{\rm s},f_{\rm PBH})}\frac{32\pi^3}{H(z_{\rm l})}\frac{D_{\rm ls}^2D_{\rm l}^2}{D_{\rm s}^2}\times\\
\frac{dn(\sigma_{\rm sis},z_{\rm l}|f_{\rm PBH})}{d\sigma_{\rm sis}}(1+z_{\rm l})^2\sigma_{\rm sis}^4y,
\end{split}
\end{equation}
where $y\in[0,1]$ is independent of other parameters and thus can be expressed as
\begin{equation}\label{eq4-11}
\begin{split}
p(y,z_{\rm l},\sigma_{\rm sis}|z_{\rm s},f_{\rm PBH})=
p(y)p(z_{\rm l},\sigma_{\rm sis}|z_{\rm s},f_{\rm PBH})\\
=\frac{2y}{\tau(z_{\rm s},f_{\rm PBH})}\frac{d\tau}{dz_{\rm l}d\sigma_{\rm sis}}.
\end{split}
\end{equation}
Therefore, $p(\Delta t|f_{\rm PBH})$ can be simplified to
\begin{equation}\label{eq4-12}
p(\Delta t|f_{\rm PBH})=\int dy\frac{p(y)}{y}\bar{p}(\Delta t/y|f_{\rm PBH}).
\end{equation}
By taking advantage of the properties of the $\delta$-function and defining $\bar{\Delta t}\equiv\Delta t/y$, $\bar{p}(\Delta t/y|f_{\rm PBH})$ is
\begin{equation}\label{eq4-13}
\begin{split}
&\bar{p}(\Delta t/y|f_{\rm PBH})=\int\sigma_{\rm sis} \int dz_{\rm l}\int dz_{\rm s}\\
&\delta(\bar{\Delta t}-\bar{\Delta t}(\sigma_{\rm sis},z_{\rm l},z_{\rm s}|f_{\rm PBH}))
p(z_{\rm l},\sigma_{\rm sis}|z_{\rm s},f_{\rm PBH})p(z_{\rm s})=\\
&\int dS\frac{p(z_{\rm l},\sigma_{\rm sis}|z_{\rm s},f_{\rm PBH})p(z_{\rm s})}{|\nabla(\bar{\Delta t}(\sigma_{\rm sis}',z_{\rm l},z_{\rm s}|f_{\rm PBH}))|},~(\sigma_{\rm sis}' = \frac{\bar{\Delta t}^{1/4}}{K(z_{\rm l}, z_{\rm s})})
\end{split}
\end{equation}
where $S$ represents the iso-time delay plane of $\bar{\Delta t}=\bar{\Delta t}(\sigma_{\rm sis},z_{\rm l},z_{\rm s}|\Omega)$, and $K(z_{\rm l}, z_{\rm s})$ is 
\begin{equation}\label{eq4-14}
K(z_{\rm l}, z_{\rm s})= \bigg(32\pi^2(1+z_{\rm l})\frac{D_{\rm l}D_{\rm ls}}{D_{\rm s}}\bigg)^{1/4}.
\end{equation}
The right panel of Figure~\ref{fig2} presents the distribution of time delays for strong lensing GW events under different cosmological scenarios. It is found that the $\Lambda$CDM+PBH cosmology, owing to the inclusion of large-scale power-spectrum contributions, produces a greater number of strong lensing events with high time delays. For sufficiently small values of the SMPBH fraction $f_{\rm PBH}\ll10^{-4}$, both the time delay distribution and the number of strong lensing events are nearly indistinguishable from those in the conventional $\Lambda$CDM scenario.

However, the finite observing time $T_{\rm obs}$ limits the number of detectable events and truncates the distribution of time delay. Hence, the detectable count of strongly lensed GW events is given by
\begin{equation}\label{eq4-15}
\Lambda_{\rm L,GW}(f_{\rm PBH},T_{\rm obs}) = \mathcal{S}(f_{\rm PBH},T_{\rm obs})\Lambda_{\rm L,GW}(f_{\rm PBH})
\end{equation}
where $\mathcal{S}(f_{\rm PBH},T_{\rm obs})$ takes into account the selection effects introduced by the finite observation time $T_{\rm obs}$
\begin{equation}\label{eq4-16}
\begin{split}
\mathcal{S}(f_{\rm PBH},T_{\rm obs})=\int p(\Delta t|f_{\rm PBH})\times\\
\frac{(T_{\rm obs}-\Delta t)}{T_{\rm obs}}\Theta(T_{\rm obs}-\Delta t) d\Delta t,
\end{split}
\end{equation}
where $\Theta(T_{\rm obs}-\Delta t)$ is Heaviside step function. As shown in Figure~\ref{fig3} for the $\Lambda$CDM case, the detectable time delay distribution $p(\Delta t|f_{\rm PBH},T_{\rm obs})$ can be derived from the expected distribution $p(\Delta t|f_{\rm PBH})$ by applying cutoff that excludes time delays longer than the observation period $T_{\rm obs}$ as
\begin{equation}\label{eq4-17}
\begin{split}
p(\Delta t|f_{\rm PBH},T_{\rm obs})= \frac{1}{Z_{\Delta t}}p(\Delta t|f_{\rm PBH})\times\\
\frac{(T_{\rm obs}-\Delta t)}{T_{\rm obs}}\Theta(T_{\rm obs}-\Delta t),
\end{split}
\end{equation}
where $Z_{\Delta t}$ is  normalization factor. 

\section{Constraints on SMPBH}\label{sec4}
These distinct signatures in the time delay distribution and the total event rate of strong lensing can be utilized to either directly infer the abundance of SMPBHs or to place a robust lower limit on their fractional contribution $f_{\rm{PBH}}$. To assess the ability of our method to constrain the abundance of SMPBHs $f_{\rm{PBH}}$, we choose $\Lambda$CDM cosmology as fiducial model. For a given set of assumptions regarding the BBH detection rate and observing time $T_{\rm obs}$, we simulate mock observations by drawing the observed number of strongly lensed events $N_{\mathrm{obs}}$ from a Poisson distribution with mean $\Lambda_{\mathrm{L,GW}}(f_{\rm PBH}, T_{\rm obs})$, and subsequently generating their time delays $\{\Delta t_i\}_{i=1}^{N{\rm{obs}}}$ according to the conditional distribution $p(\Delta t|f_{\rm PBH}, T_{\rm obs})$. As shown in Figure~\ref{fig3}, the upper panel presents the detectable time delay distributions and event counts for a fixed observational duration $T_{\rm{obs}} = 10~\rm{yrs}$ under four BBH detection rates: $R\in[5 \times 10^5,\; 1 \times 10^5,\; 5 \times 10^4,\; 1 \times 10^4]~\rm{yr^{-1}}$. The lower panel follows a similar analysis but with the detection rate fixed at $R = 1 \times 10^5~\rm{yr^{-1}}$ and for varying observational durations $T_{\rm{obs}}\in[10,\;5,\;2,\;1]~\rm{yrs}$. The comparison between the theoretical and simulated cumulative distribution functions (CDFs) yields a K-S statistic of $D<0.1$, which indicates that the simulated data are consistent with the predictions of the $\Lambda$CDM cosmology.

We consider above scenario where $N_{\rm obs}$ strong lensing BBH events, each producing two observable images, are confidently detected within an observing period $T_{\mathrm{obs}}$. Using the set of measured time delays $\{\Delta t_i\}_{i=1}^{N{\rm{obs}}}$ from these events, we compute the posterior distribution for the abundance of SMPBHs $f_{\rm{PBH}}$ within the $\Lambda$CDM+PBH cosmology, adopting a log-uniform prior. The likelihood $p(\{\Delta t_i\}_{i=1}^{N{\rm{obs}}}|f_{\rm PBH},T_{\rm obs})$ in the hierarchical Bayesian inference framework is 
\begin{equation}\label{eq5-1}
\begin{split}
&p(\{\Delta t_i\}_{i=1}^{N{\rm{obs}}}|f_{\rm PBH},T_{\rm obs})=\prod_{i}^{N_{\rm obs}}p(\Delta t_{i}|f_{\rm PBH},T_{\rm obs})\times\\
&\frac{[\Lambda_{\rm L,GW}(f_{\rm PBH},T_{\rm obs})]^{N_{\rm obs}}e^{-\Lambda_{\rm L,GW}(f_{\rm PBH},T_{\rm obs})}}{N_{\rm obs}!}.
\end{split}
\end{equation}
In this analysis, it is assumed that each BBH merger constitutes an independent event, and that the associated time delays are determined with high accuracy and precision.

Then, we estimate the upper limit of $f_{\rm{PBH}}$ in the $\Lambda$CDM+PBH cosmology with $M_{\rm{PBH}} = 10^9~M_{\odot}$ by incorporating the simulated lensing time delay sample $\{\Delta t_i\}_{i=1}^{N{\rm{obs}}}$ into the \textbf{EMCEE} package~\cite{emcee} using the posterior distribution given by Equation~(\ref{eq5-1}).
Figure~\ref{fig4} shows the $95\%$ quantile of the posterior distribution for four scenarios: 1) for initially Poisson distributed SMPBHs ($\xi_0 = 0$, upper left panel), the $95\%$ upper limit on $\log(f_{\rm{PBH}})$ decreases from $-3.76$ to $-3.19$ as the BBH detection rate $R$ is reduced; 2) for the same Poisson distributed SMPBHs (upper right panel), the limit decreases from $-3.46$ to $-3.06$ as the observational duration $T_{\rm{obs}}$ is shortened; 3) for initially clustered SMPBHs ($x_{\rm{cl}} = 1~\rm{Mpc}$, $\xi_0 = 10$, lower left panel), the limit decreases from $-3.94$ to $-3.53$ with decreasing $R$; 4) for the same clustered SMPBHs (lower right panel), the limit decreases from $-3.76$ to $-3.34$ with shorter $T_{\rm{obs}}$. From these results, two key trends are evident: firstly, increasing the number of strong lensing GW events leads to more stringent constraints on $f_{\rm{PBH}}$; secondly, compared to the Poisson distributed model, the clustered model induces a stronger modification to the power spectrum, which in turn yields more stringent upper limits on $f_{\rm{PBH}}$. 

\begin{figure}
    \centering
    \includegraphics[width=0.49\textwidth, height=0.35\textwidth]{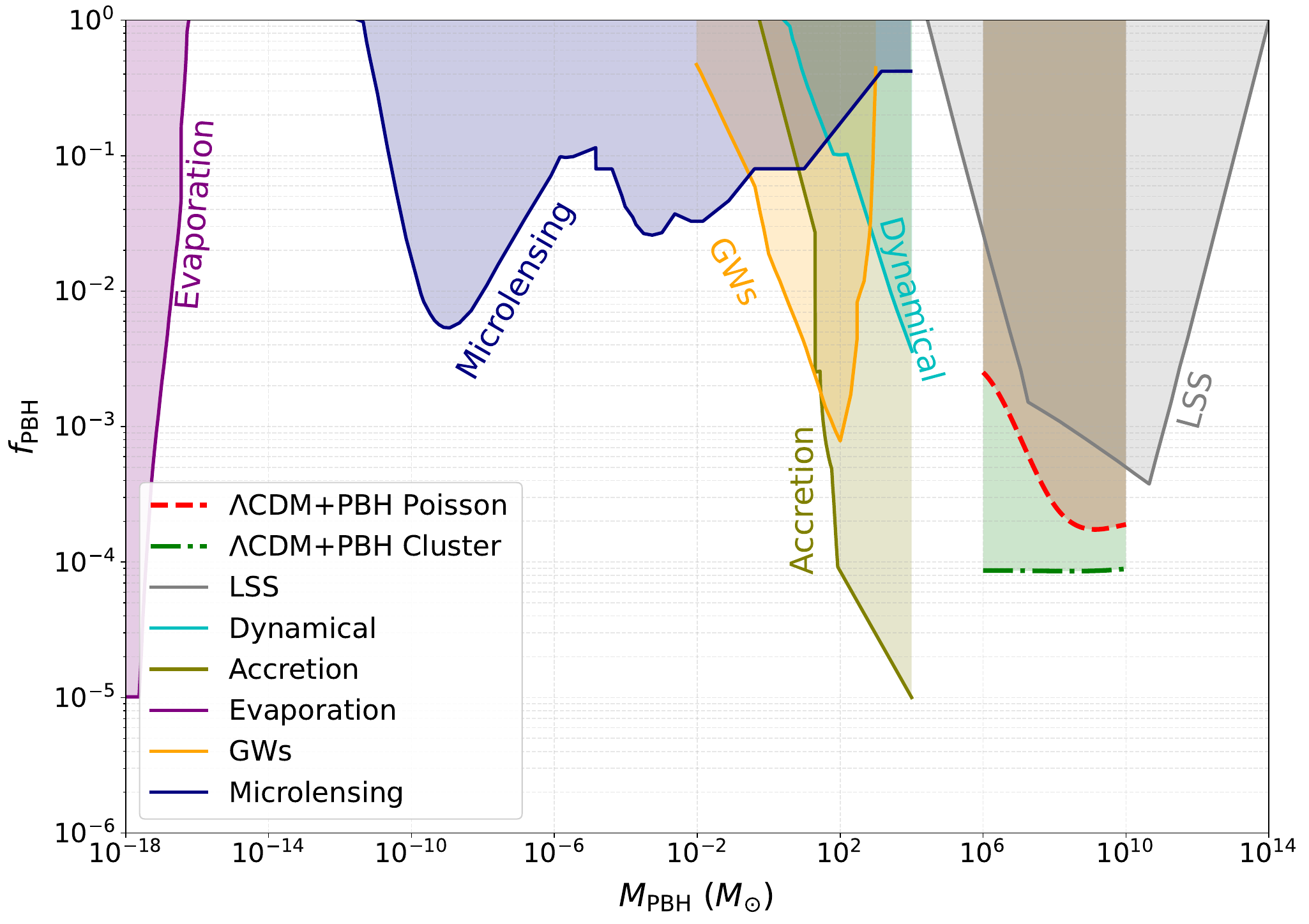}
     \caption{The upper limit of $f_{\rm PBH}$ at $95\%$ confidence level versus SMPBH mass $M_{\rm PBH}\in[10^6,10^{10}]~M_{\odot}$ for fixed BBH detection rates $R=5\times10^5~{\rm yr^{-1}}$ and observational durations $T_{\rm obs}=10~{\rm yrs}$. The red dashed and green dash-dotted curves denote the $\Lambda$CDM+PBH cosmology with initially Poisson distributed ($\xi_0=0$) and clustered SMPBHs ($x_{\rm cl}=10~{\rm Mpc}$, $\xi_0=10$), respectively. Other constraints are compiled from existing reviews~\cite{Green:2020jor,Carr:2020gox,Carr:2021bzv}: these include limits from Hawking radiation evaporation (Evaporation), micro-lensing surveys (Microlensing), Stochastic GW background (GWs), effect of accretion (Accretion), dynamical effects (Dynamical), and the imprint of PBHs on large-scale structure (LSS).
     }\label{fig5}
\end{figure}

As shown in Figure~\ref{fig5}, the $95\%$ confidence level upper limit on $f_{\rm PBH}$ is plotted against the SMPBH mass $M_{\rm PBH}\in [10^6, 10^{10}]~M_{\odot}$, for fixed BBH detection rate $R = 5 \times 10^5~{\rm yr^{-1}}$ and observational duration $T_{\rm obs} = 10~{\rm yrs}$ in the $\Lambda$CDM+PBH cosmology. For the Poisson distributed model, the upper limit on $\log(f_{\mathrm{PBH}})$ tightens from $-2.60$ to $-3.72$ as SMPBH mass $M_{\mathrm{PBH}}$ increases, since the isotropic power spectrum $P_{\rm iso, Poisson}$ is proportional to $f_{\rm PBH}M_{\rm PBH}$. In the clustered model ($x_{\rm cl}=10~{\rm Mpc}$, $\xi_0=10$), where $P_{\rm iso,Cluster}\propto f_{\rm PBH}^2\xi_0 x_{\rm cl}^3 \gg P_{\rm iso,Poisson}$, the total isotropic power spectrum $P_{\rm iso}$ is dominated by the clustered component and thus becomes independent of SMPBH mass $M_{\rm PBH}$, resulting in the upper limit on $\log(f_{\rm PBH})\simeq-4.06$ that is both more stringent and independent of $M_{\rm PBH}$. Compared with other constraints in Figure~\ref{fig5}, the precisely measured time delays of strong lensing GWs would provide a powerful probe for constraining SMPBHs.

\section{Conclusion and Discussion}\label{sec5}
Precise time delay measurements from strongly lensed GW events, expected to be detected in great numbers by next-generation ground-based GW detectors, could provide complementary constraints on the nature of dark matter. A promising candidate for this component within the $\Lambda$CDM framework is a population of SMPBHs. Such a component could enhance the early formation of dark matter halos, potentially accounting for the abundance of high-redshift galaxies observed by the JWST. Importantly, the $\Lambda$CDM+PBH cosmology predicts a unique signature in the population of strongly lensed GWs, i.e. an overall higher rate of such events, accompanied by a characteristically altered distribution of time delays between multiple images. Therefore, we propose a method to directly constrain the abundance of SMPBHs in dark matter $f_{\rm PBH}$ from the observed distribution of time delays and the event rate of strongly lensed GWs from future detections. As illustrated in Figure~\ref{fig5}, the constraint derived from our lensed GWs analysis ($f_{\rm PBH}\sim10^{-4}$ for $M_{\rm PBH}>10^8 M_\odot$ at $95\%$ confidence level) is complementary to bounds from other probes and even more stringent in certain regimes. 

While the method we propose constitutes a powerful new probe for constraining SMPBHs, it is important to note that the present analysis is based on a relatively ideal scenario. The inclusion of key limitations currently absent from our modeling, i.e., selection effects and measurement uncertainty caused by systematic error, would improve the precision of our constraints. These factors, as encapsulated in Equation~(\ref{eq5-1}) and detailed below, including:
\begin{itemize}
\item \textbf{Detection Threshold and Duty Cycle}:
The statistical inference from strongly lensed GW events depends critically on observational completeness. Key factors affecting this completeness include: (i) whether the signal-to-noise ratio of secondary images meets the detection threshold, and (ii) the non-continuous duty cycle of the GW detector network. While population properties (e.g., the BBH merger redshift distribution) can be independently informed by unlensed signals, a robust strong lensing analysis must explicitly correct for these coupled selection effects to avoid biases in the inferred event rates and time delay distributions.

\item \textbf{Lens Model Selection}:
Our analysis adopts a simplified lens model, describing lenses as SIS whose parameters are derived from the halo mass function using a straightforward prescription. This treatment does not account for halo substructure or baryonic effects. Consequently, the choice of lens model may introduce potential biases that couple both selection effects and measurement uncertainties. Future work should therefore incorporate more realistic and physically detailed lens models.

\item \textbf{SMPBH Model Complexity}:
Our analysis adopts simplified prescriptions for two physical aspects of the SMPBH scenario: (i) the omission of local “seed” effect of individual black holes on their surrounding density field~\cite{Carr:2018rid}, and (ii) a simplified model for clustered SMPBHs. Implementing more complex, physically motivated models for both effects is necessary, as current simplifications can jointly bias the detectability of lensed events (selection effect).
\end{itemize}
Therefore, by continuing to refine this approach and combining the unprecedented sensitivity of next-generation ground-based GW detectors with the deep optical galaxy catalogs from new-generation wide-field surveys such as the Large Synoptic Survey Telescope (LSST)~\cite{LSST:2008ijt} and the China Space Station Telescope (CSST)~\cite{CSST:2025ssq}, strongly lensed GW events will be established as a high-precision cosmological probe.

\section{Acknowledgements}
This work is supported by National Key R\&D Program of China under Grant No.2024YFC2207400; Research Performance Assesssment Grant of the Postdoctoral Fellowship Program of China Postdoctoral Science Foundation under Grant No.YJB20250367; National Natural Science Foundation of China under Grants Nos.12322301, 12275021, 12503002, and 12447137.

\bibliographystyle{apsrev4-1}
\bibliography{ref}
\end{document}